# Stochastic inner workings of subdiffraction laser writing

JULIA M. MIKHAILOVA,[1] ALEKSEI M. ZHELTIKOV[2,*]

[1] *Department of Mechanical and Aerospace Engineering, Princeton University, Princeton, New Jersey, 08544*
[2] *IQSE, Department of Physics and Astronomy, Texas A&M University, College Station TX 77843*
*\*zheltikov@physics.tamu.edu*

**Ultrafast laser writing of single lattice defects in wide-bandgap semiconductors is shown to present a new physical setting in which deeply subwavelength laser-writing positioning precision is attainable, but where the whole notion of positioning can only be understood in a statistical sense. We outline a framework for the analysis of this class of laser – matter interactions, grounding the concepts of optical super-resolution and subdiffraction positioning in statistical optics. Working along these lines, we derive closed-form solutions for physically meaningful quantifiers of laser–matter interactions on a subwavelength scale, suggesting a physically clear view of how deeply subdiffraction resolution can emerge from the interplay between determinism and stochasticity. We show that subdiffraction positioning precision in single-lattice-defect laser writing is achieved at the cost of a lower throughput, setting physical bounds on the scalability of integrated quantum photonic systems fabricated by means of super-resolving laser writing.**

Modern optical science and advanced laser technologies provide powerful resources whereby fundamental processes in physical, chemical, and biological systems can be probed and tailored with an unprecedented temporal [1, 2], spatial [3], and spectral [4] resolution. Latest achievements in ultrafast optics enable time-resolved studies on the attosecond time scale and help establish an ultimate control over electron dynamics in a vast variety of atomic, molecular, and solid-state systems [1, 2]. High-precision positioning of the sites of laser-driven excitations and laser-energy deposition, on the other hand, is the key for resolving an ultrafine motion of atoms in molecules in field – matter interactions [1, 5] and is central for the advancement of laser-material-processing technologies [6, 7]. Leading the frontier of this research are the efforts to develop methods of high- precision laser writing in solid-state materials as a cutting-edge approach to laser material processing [8 – 11]. As a recent milestone achievement, methods of femtosecond laser writing have been shown to enable high-precision positioning of single spin-defect centers in wide-bandgap materials, such as diamond [12 – 16], silicon carbide (SiC) [17 – 21], hexagonal boron nitride [22 – 25], and aluminum nitride [26], thus opening routes toward spin-center-based scalable all-solid integrated quantum photonics and quantum-information architectures.

Although the laser writing of single spin-defect centers in wide-bandgap materials has been, on many occasions, viewed as deterministic, comprehensive quantitative studies [17, 18] reveal clear signatures of stochasticity in single-spin-defect laser writing. These findings raise fundamental questions regarding the role of stochasticity in deeply subdiffraction optical modalities, such as those activated in single-defect laser writing, leading us in many cases to rethink and redefine the notions of resolution and precision, seeking the ways to meaningfully connect these concepts to statistical optics. Finding an adequate framework to address these questions is key to understanding how deeply subdiffraction resolution can emerge from the interplay between determinism and stochasticity and is central for understanding the ultimate capabilities and fundamental limitations of rapidly growing technologies of laser material processing, including single-center laser writing in solids [14, 17, 18, 25].

The purpose of this work is to lay the groundwork for the development of such a framework. To this end, we consider an ultrashort laser pulse with a central frequency $\omega_0$, pulse width $\tau_0$, and a field-intensity beam profile $I(r)$, interacting with a transparent solid semiconductor/dielectric with a band gap $E_g$ and refractive index $\eta_0$. As the laser field gives rise to ultrafast electron transitions from bonding states within the valence band to antibonding states in the conduction band, via multiphoton or tunneling pathways, interatomic bonds within the lattice are softened, facilitating random displacements of atoms from their lattice sites [27 – 36]. In real space, this process is understood in terms of laser-driven charge redistribution, whereby antibonding charges tend to spread along interatomic bonds, inducing dynamic stretching forces on these bonds. In sufficiently intense laser fields, this process gives rise to bond cleavage [31 – 38], leading to a transition from crystalline solid to a disordered liquid at lattice temperatures well below the melting point and on time scales much shorter than the times required for the lattice to reach its melting point.

Serving as a key quantifier for such ultrafast lattice dynamics is the mean-square displacement $\langle [u(t)]^2 \rangle$ of atoms relative to their equilibrium sites in the crystal lattice. Because the dynamics of $\langle [u(t)]^2 \rangle$ is readily accessible via time-resolved x-ray diffraction [31 – 33], connecting to the directly measurable x-ray diffraction intensity, $I(q,t) = \exp\{-q\langle [u(t)]^2\rangle/3\}$, $q$ being the reciprocal lattice constant, many successful models of ultrafast laser – solid interactions operate with the Lindemann criterion [39 – 42], identifying the loss of stability within a crystal lattice in terms of the root-mean-square (RMS) displacement $\sqrt{\langle [u(t)]^2 \rangle}$ and placing the empirical threshold for such an instability at $u_s \approx 0.15d$, i.e., at $\approx 15\%$ of the nearest-neighbor distance $d$ within the crystal lattice. The applicability realm of this approach spans across a broad variety of laser – solid interaction regimes and technologically significant semiconductors [27 – 38].

Given that the probability density function (PDF) of atomic displacements $u$ is $w(u)$, the probability that an individual atom in the laser-irradiated crystal lattice is displaced from its lattice site by more than $u$ is

$$P(u) = \int_u^\infty w(u) du. \qquad (1)$$

Specifically, with $u$ identified with the threshold $u_s$, $P(u = u_s)$ gives the probability that an individual atomic displacement gives rise to a lattice defect. In a laser beam with field intensity profile $I(r)$, both the PDF $w(u)$ and its integral transform $P(u)$ are functions of $r$, $w(u) = w(u,r)$ and $P(u) = P(u,r)$. The expected number of laser-induced defect centers within a laser-irradiated crystal area with a surface density of lattice sites $n_0$ is then found as

$$q = 2\pi \int_0^\infty n(r) r dr, \qquad (2)$$

where $n(r) = n_0 P(u_s, r) = 2\pi n_0 \int_{u_s}^\infty w(u,r) du$ is the spatial distribution of laser-induced defects.

The requirement that, in a statistical sense, the laser driver induces one defect center within the laser-irradiated crystal area can thus be expressed as

$$q = 2\pi n_0 \int_0^\infty r dr \int_{u_s}^\infty w(u,r) du = 1. \qquad (3)$$

While Eq. (3) guarantees that, statistically, the laser driver induces one defect center in a laser-irradiated area within a crystal sample, the position of this lone defect within the laser-irradiated area is random and can only be described in probabilistic terms. To define these terms, we consider normally distributed atomic displacements, with

$$w(x) = (2\pi)^{-1/2} \exp(-x^2/2), \quad x = u/\sqrt{\langle u^2\rangle}. \qquad (4)$$

With $w(x)$ as defined by Eq. (4) and with the lower integration limit $u$ in Eq. (1) identified with $u_s$, we find $P(u_s) = (1/2)\mathrm{erfc}\left(u_s/\sqrt{2\langle u^2\rangle}\right)$, where $\mathrm{erfc}(x)$ is the complementary error function of $x$. Because the number of lattice sites irradiated by a laser beam is very large, the regime of single-defect generation is only attainable when $u > u_s$ exceedances are rare, i.e., $u_s \gg \sqrt{\langle u^2\rangle}$. In this limit, $x \gg 1$, $\mathrm{erfc}(x/\sqrt{2}) \approx \sqrt{2/\pi}\, x^{-1} \exp(-x^2/2)$, leading to

$$n(r) \approx (n_0/\sqrt{2\pi}) \left\{\sqrt{\langle [u(r)]^2\rangle}/u_s\right\} \exp[-u_s^2/(2\langle [u(r)]^2\rangle)]. \qquad (5)$$

To gain insights into the properties of the variance $\langle u^2\rangle$ as a function of the laser intensity and the transverse coordinate $r$, we consider an archetypal setting of ultrafast laser writing, in which single defect centers are induced in a wide-bandgap semiconductor with laser pulses whose photon energy $\hbar\omega_0$ is lower than the bandgap $E_g$ and whose field intensity drives interband electron transitions via nonlinear photoionization. Specifically, in a rapidly growing technology [17 – 21] of laser writing of single spin centers in SiC ($E_g \approx 3.3$ eV, $\eta_0 \approx 2.6$), well-isolated single lattice defects are induced by ultrashort Ti: sapphire-laser pulses with a central wavelength $\lambda_0 \approx 800$ nm, corresponding to the photon energy $\hbar\omega_0 \approx 1.55$ eV, typical pulse width $\tau_0 \approx 250$ fs, and field intensities at the level of $I \approx 10^{13}$ W/cm$^2$. At this level of $I$, the Keldysh photoionization parameter for $\hbar\omega_0 \approx 1.55$ eV, $E_g \approx 3.3$ eV, $\eta_0 \approx 2.6$, and the transverse effective electron mass $m_\perp \approx 0.42 m_e$ [43, 44] is above the $\gamma_K = 1$ borderline between multiphoton and tunneling ionization. Laser-driven interband electron dynamics in this regime is dominated by $n$-photon ionization with $n = \lceil E_g/(\hbar\omega_0) \rceil = 3$, leading to

$$\sqrt{\langle [u(r)]^2\rangle} \approx \varepsilon [I(r)]^n, \qquad (6)$$

with $\varepsilon = \kappa_n \tau_0$ and $n$-photon ionization coefficient $\kappa_n$.

For a laser driver with a Gaussian beam profile, $I(r) = I_0 \exp(-r^2/r_0^2)$, Eqs. (3) – (6) combine into

$$n(r) \approx (n_0/\sqrt{2\pi})(\delta_0/u_s) \exp(-nr^2/r_0^2)$$
$$\exp[-u_s^2/(2\delta_0^2) \exp(2nr^2/r_0^2)], \qquad (7)$$

with $\delta_0^2 = \varepsilon^2 I_0^{2n} = \langle [u(r=0)]^2\rangle$.

With $\exp(2nr^2/r_0^2)$ approximated with its power-series expansion about $r = 0$, Eq. (7) yields

$$n(r) \approx (n_0/\sqrt{2\pi})(\delta_0/u_s) \exp[-u_s^2/(2\delta_0^2)] \exp(-\alpha r^2), (8)$$
$$\alpha = nr_0^{-2}(1 + u_s^2/\delta_0^2). \qquad (9)$$

Eq. (2) then gives

$$q \approx (\pi/2)^{1/2} (n_0/\alpha)(\delta_0/u_s) \exp[-u_s^2/(2\delta_0^2)]. \qquad (10)$$

Provided that the laser driver induces $k = 1$ lattice defect within the entire laser-irradiated area, $q = 2\pi \int_0^\infty n(r) r dr = 1$, the probability that this sole defect is found within a circle of radius $r_c$ around the beam center $r = 0$ is

$$\wp(r < r_c | k = 1) \approx 1 - \exp(-\alpha r_c^2). \qquad (11)$$

For small $r_c$, such that $r_c \ll 1/\sqrt{\alpha}$, Eq. (11) gives

$$\wp(r < r_c | k = 1) \approx \alpha r_c^2 = n(1 + u_s^2/\delta_0^2)(r_c/r_0)^2. \qquad (12)$$

For large $r_c$, on the other hand, such that $r_c \gg 1/\sqrt{\alpha}$, $\wp(r < r_c | k = 1) \to 1$. The $k = 1$ conditional probability that the sole defect induced within the laser beam falls outside a circle of radius $r_c$ around $r = 0$ is then exponentially small, $1 - \wp(r < r_c | k = 1) \approx \exp(-\alpha r_c^2) \ll 1$.

To gain better insights into the physical content of these findings, we express the distribution of laser-induced defects $n(r)$ found in Eq. (7) as

$$n(r) \approx (n_0/\sqrt{2\pi})(\delta_0/u_s) \exp[-u_s^2/(2\delta_0^2)] \exp(-r^2/\rho_0^2), (13)$$

with $\rho_0 = 1/\sqrt{\alpha} = r_0/\sqrt{n(1 + u_s^2/\delta_0^2)}$.

It is readily seen from Eq. (13) that the parameter $\rho_0$ serves as a measure of spatial confinement of laser-induced lattice-defect generation. It is instructive to express this parameter as a product

$$\rho_0 = r_0 n^{-1/2}(1 + u_s^2/\delta_0^2)^{-1/2} = r_0/(c_1 c_2) \qquad (14)$$

of the laser-beam radius $r_0$ and two confinement-enhancement ratios, $c_1 = \sqrt{n}$ and $c_2 = \sqrt{1 + u_s^2/\delta_0^2}$.

$c_1$ is readily recognized as a factor that quantifies confinement enhancement due to the $n$ th-order optical nonlinearity – effect that is widely employed in $n$-photon microscopy [45 – 49]. The width of the spatial profile of the photoelectron density $n_e(r)$ induced via $n$-photon ionization by a Gaussian beam with a beam radius $r_0$ is $r_0/\sqrt{n}$. The probability of atomic displacement, on the other hand, is a nonlinear function of the photoionization rate. Due to this nonlinearity, the spatial confinement of the distribution $n(r)$ of laser-induced defects within the laser beam is even tighter than the confinement of the photoelectron density $n_e(r)$. The factor that quantifies such enhancement is $\sqrt{1 + u_s^2/\delta_0^2}$, i.e., the third multiplier in Eq. (14). The spatial confinement $\rho_0$ of laser-induced defect distribution $n(r)$ is thus a product of the laser beam radius $r_0$ and two factors, $c_1$ and $c_2$, expressing confinement enhancement due to, respectively, the nonlinearity of $n$-photon ionization and nonlinearity of the $n_e(r)$-to-$n(r)$ map (Table 1).

**Table 1. Stochastic properties of subdiffraction laser writing**

| Spatial distribution | $I(r)$ | $n_e(r)$ | $n(r)$ |
|---|---|---|---|
| Localization scale | $r_0$ | $r_0/\sqrt{n}$ | $r_0/\sqrt{n(1+u_s^2/\delta_0^2)}$ |
| | SiC: $E_g \approx 3.3$ eV, $\eta_0 \approx 2.6$, $d \approx 0.2$ nm | | |
| | Laser driver: $\hbar\omega_0 \approx 1.55$ eV, $\tau_0 \approx 250$ fs, $I \approx 10^{13}$ W/cm$^2$, $r_0 \approx \lambda_0$ | | |
| | $\lambda_0$ | $\lambda_0/\sqrt{3}$ | $0.12\lambda_0$ |
| $\wp(r<r_c\|k=1)$ | $1 - \exp(-\alpha r_c^2), \alpha = n r_0^{-2}(1+u_s^2/\delta_0^2)$ | | |
| | $r_c \ll 1/\sqrt{\alpha}$ | | $r_c \gg 1/\sqrt{\alpha}$ |
| | $n(1+u_s^2/\delta_0^2)(r_c/r_0)^2$ | | $\wp(r<r_c\|k=1) \to 1$ |
| $\wp_k$ | $(q^k/k!)\exp(-q)$ | | |
| $\vartheta(r_c)$ | $\wp(r<r_c\|k=1)\wp_1$ | | |
| $\vartheta_{max}(r_c)$ | $e^{-1}[1-\exp(-\alpha r_c^2)]$ | | |
| $\vartheta_m$ | $e^{-m}\{1-\exp[-n(1+u_s^2/\delta_0^2)\sigma_d^2/r_0^2]\}^m$ | | |
| | $\sigma_d = \rho_0, k=1$ | | $\sigma_d \approx \lambda_0, k=1$ |
| | $e^{-m}(1-e^{-1})^m$ | | $e^{-m}$ |

Seeking a representative estimate for the spatial confinement attainable in ultrafast laser writing, we consider the crystal lattice of SiC ($E_g \approx 3.3$ eV, $d \approx 0.2$ nm) irradiated with ultrashort pulses of a Ti: sapphire laser ($\hbar\omega_0 \approx 1.55$ eV), focused into a beam waist with $r_0$ close to the laser wavelength $\lambda_0$. We now resort to Eq. (3) as a resource to define the parameters of the laser driver needed for single-defect laser writing. Using the solution for $q$ found in Eq. (10) and solving Eq. (3) for $u_s/\delta_0$ with laser and material parameters as defined above and with $n_0 \approx d^{-2}$, we find $c_1 \approx 1.73$, $u_s^2/\delta_0^2 \approx 22$, and $c_2 \approx 4.8$, leading to $\rho_0 \approx r_0/8.3 \approx 0.12 r_0$ (Table 1). Nonlinearity of laser-induced lattice-defect generation in a wide-bandgap semiconductor is thus seen to significantly enhance laser-field confinement, enabling ultrafast laser writing of single defect centers with a subdiffraction positioning precision.

A steeper gradient of the electron density $n_e(r)$ gives rise to a faster electron diffusion, $\partial n_e/\partial t = D_e \partial^2 n_e/\partial x^2$. With the room-temperature electron diffusion coefficient of SiC ranging from $D_e \approx 20$ cm$^2$/s for 3C-SiC to $D_e \approx 90$ cm$^2$/s for 6H-SiC [50], $n$-photon ionization of SiC by a $\lambda_0 \approx 800$ nm laser beam focused into a spot with a typical size of $\lambda_0/\eta_0$ gives rise to an electron density gradient $\ell_e = n_e^{-1}|\partial^2 n_e/\partial x^2| \approx \lambda_0/(\eta_0\sqrt{n})$. With $\eta_0 \approx 2.6$, $n = \lceil E_g/(\hbar\omega_0)\rceil = 3$, and $D_e \approx 20$ cm$^2$/s, such a gradient induces electron diffusion on a typical time scale $\tau_d \approx \ell_e^2/D_e \approx 20$ ps. This estimate shows that, even with electron density gradients induced via multiphoton ionization on a deeply subwavelength scale, $\lambda_0/(\eta_0\sqrt{n}) \approx 0.2\lambda_0$ for SiC, the typical time of electron diffusion is still significantly longer than the laser pulse widths $\tau_0$ in ultrafast laser writing.

Because Eq. (3) can guarantee that $k = 1$ only in a statistical sense, we now need to complete the statistical analysis of single-defect laser writing by finding the probability that $k = 1$ provided that the laser parameters are adjusted in such a way as to satisfy Eq. (3) with $q$ as defined in Eq. (10). To this end, we combine Eqs. (7) – (9) to express the expected number of laser-induced lattice defects within the laser-irradiated area as

$$q = 2\pi n_0 q_0 \int_0^\infty \exp(-\alpha r^2) \, r dr, \quad (15)$$

with $q_0 = (2\pi)^{-1/2}(\delta_0/u_s)\exp[-u_s^2/(2\delta_0^2)]$ and $\alpha$ as found in Eq. (9).

Integration in Eq. (15) is straightforward, yielding

$$q = \pi r_0^2 n_0 (1 + u_s^2/\delta_0^2)^{-1} q_0/n. \quad (16)$$

Recognizing $\mathcal{N} = \pi r_0^2 n_0 n^{-1}(1 + u_s^2/\delta_0^2)^{-1}$ as the number of particles within a circle of radius $\rho_0 = 1/\sqrt{\alpha} = r_0/\sqrt{n(1+u_s^2/\delta_0^2)}$, we find that the expected number of laser-induced lattice defects within the laser-irradiated area factorizes as a product $q = \mathcal{N} q_0$ of the number of particles within a circle of radius $\rho_0$ and $q_0$, which can be viewed as the effective probability of laser-induced defect generation per lattice site within the laser-irradiated area. The probability that defect centers are generated in $k$ out of $\mathcal{N} = N$ such lattice sites is $\mathcal{P}_N^k = C_N^k [q_0]^k [1-q_0]^{N-k}$, with $C_N^k = N!/[k!(N-k)!]$. Expressing $q_0$ in $\mathcal{P}_N^k$ as $q_0 = q/N$ and observing that $\lim_{N\to\infty} N!/[N^k(N-k)!] = 1$, $\lim_{N\to\infty}[1-q/N]^{-k} = 1$, and $\lim_{N\to\infty}[1-q/N]^N = \exp(q)$, we find

$$\wp_k = \lim_{N\to\infty} \mathcal{P}_N^k = (q^k/k!)\exp(-q). \quad (17)$$

$\wp_k$ in Eq. (17) is readily recognized as the Poisson distribution. This result is fully consistent with the experimental studies of the statistical properties of laser-induced color-center generation in SiC [17, 18], reporting, among other important findings, the Poisson statistics of color centers in a laser-irradiated area.

Eqs. (7) – (14), (17) provide a fully operation framework for the analysis of the statistical properties of laser-driven lattice-defect generation. The task of single-defect generation can now be understood and quantitatively described in probabilistic terms. Specifically, the probability that the number of lattice defects in the laser irradiated area is one is found by setting $k = 1$ in Eq. (17), leading to

$$\wp_1 = \wp_1(u) = q(u)\exp[-q(u)]. \qquad (18)$$

Regardless of the specific form of $q(u)$ as a function of $u$, the maximum of $\wp_1$, $\wp_1^{\max} = e^{-1}$, is achieved when $q(u) = 1$, i.e., $\langle k \rangle = 1$. Thus, a laser writing experiment repeated $S$ times, e.g., on $S \gg 1$ identical samples, or, at $S \gg 1$ identical sites within one sample [17 – 25], will yield $\wp_1 S$ samples with $k = 1$ lattice defect in each. For $\wp(r < r_c | k = 1)\wp_1 S$ samples within this subset, this sole defect will be found within a circle of radius $r_c$ around the beam center. The product $\vartheta(r_c) \approx \wp(r < r_c | k = 1)\wp_1$ is thus identified as the throughput of single-center laser writing with positioning precision $r_c$. With laser parameters adjusted in such a way as to satisfy Eq. (3), yielding, on the average, one defect center within the laser-irradiated area, the throughput $\vartheta$ is maximized at $\vartheta_{\max}(r_c) \approx e^{-1}\wp(r < r_c | k = 1)$.

We now consider a process aimed at writing two isolated lattice defects separated by a distance $d > \lambda_0$, each positioned with a precision $\sigma_d < \lambda_0$. This task is accomplished by laser writing with two identical laser beams whose centers are separated by $d$. When repeated $S$ times, i.e., on $S \gg 1$ identical samples, or, at $S \gg 1$ identical sites within one sample, such a two-beam writing process yields $[\vartheta(r_c = \sigma_d)]^2 S$ samples with two isolated lattice defects in which each of the two defects is located within a circle of radius $r_c = \sigma_d$ around the center of the respective laser beam. The distance between these defects is thus $d \pm \sigma_d$, with $d > \lambda_0$ and $\sigma_d < \lambda_0$. Setting the desired positioning precision at $\sigma_d = \rho_0 = r_0/\sqrt{n(1 + u_s^2/\delta_0^2)}$ and adjusting the laser parameters in such a way as to satisfy Eq. (3), we find $\vartheta_2 \approx e^{-2}(1 - e^{-1})^2$. The scalability in subwavelength-positioning single-center laser writing thus comes at a cost of lower throughput (Table 1). Specifically, the throughput of $m$-beam laser writing of an array of $m$ isolated lattice defects in which the distance between any of two centers is larger than $\lambda_0$ and in which each defect is positioned with a precision $\sigma_d < \lambda_0$, is $[\vartheta(r_c = \sigma_d)]^m$. With $\sigma_d = \rho_0$ and with laser parameters adjusted to meet Eq. (3), the throughput of this process becomes $\vartheta_m \approx e^{-m}(1 - e^{-1})^m$. When the desired defect positioning precision is close to the laser wavelength, $\sigma_d \approx \lambda_0$, on the other hand, Eq. (11) dictates $1 - \wp(r < r_c \approx \lambda_0 | k = 1) \ll 1$. For laser writing in SiC with $\hbar\omega_0 \approx 1.55$ eV, $\tau_0 \approx 250$ fs, $I \approx 10^{13}$ W/cm$^2$, and $r_0 \approx \lambda_0$, as in the example above, Eqs. (11), (14) give $1 - \wp(r < r_c \approx \lambda_0 | k = 1) \approx \exp(-c_1^2 c_2^2) \approx 10^{-30}$. The throughput of $m$-center laser writing in this setting reaches its upper bound at $\vartheta_m \approx e^{-m}$.

To summarize, ultrafast laser writing of single lattice defects in wide-bandgap semiconductors is identified as a new physical setting in subdiffraction optical physics in which deeply subwavelength positioning precision is attainable, with the notion of positioning understood in a statistical sense. We have presented an outline of a framework for the analysis of this class of laser – matter interactions and derived closed-form solutions for physically meaningful quantifiers of laser – matter interactions on a deeply subwavelength scale. Our analysis shows that subdiffraction positioning precision in single-lattice-defect laser writing is achieved at the cost of a lower throughput, setting physical bounds on the scalability of integrated quantum photonic systems fabricated by means of super-resolving laser writing.

**Funding**. DOE grant # DE-SC0024882; Welch Foundation project A-1801-20210327.

**Disclosures**. The authors declare no conflicts of interest.

**Data Availability Statement**. Data underlying the results of this paper may be obtained from the author upon request.